\documentclass[utf8]{frontiersFPHY} 

\usepackage{url,hyperref,lineno,microtype,subcaption}
\usepackage[onehalfspacing]{setspace}
\usepackage[T1]{fontenc} 
\usepackage{color}

\def\keyFont{\fontsize{8}{11}\helveticabold }
\def\firstAuthorLast{Kreisel {et~al.}} 
\def\Authors{Andreas Kreisel\,$^{1,*}$, P. J. Hirschfeld\,$^{2}$ and Brian M. Andersen\,$^{3}$}
\def\Address{$^{1}$Institut f\" ur Theoretische Physik, Universit\"at Leipzig, D-04103 Leipzig, Germany \\
$^{2}$Department of Physics, University of Florida, Gainesville, Florida 32611, USA \\
$^{3}$Niels Bohr Institute, University of Copenhagen, R\aa dmandsgade 62, DK-2200 Copenhagen, Denmark}
\def\corrAuthor{Andreas Kreisel}

\def\corrEmail{kreisel@itp.uni-leipzig.de}

\newcommand{\mbf}{\mathbf} 
\renewcommand{\k}{{\mbf k}}

\newcommand{\q}{{\mbf q}}
\def\s{\sigma}

\definecolor{blue}{rgb}{  0,  0,    1}
\definecolor{d4blue}{rgb}{  0,  0.4470,    0.7410}
\definecolor{d12orange}{rgb}{0.8500    0.3250    0.0980}
\definecolor{g8yellow}{rgb}{0.    0.6    0.298}
\definecolor{ppurple}{rgb}{0.4940    0.1840    0.5560}

\begin{document}
\onecolumn
\firstpage{1}

\title[]{Theory of spin-excitation anisotropy in the nematic phase of FeSe obtained from RIXS measurements} 

\author[\firstAuthorLast ]{\Authors} 
\address{} 
\correspondance{} 

\extraAuth{}

\maketitle

\begin{abstract}

Recent resonant inelastic x-ray scattering (RIXS) experiments have detected a significant high-energy spin-excitation anisotropy in the nematic phase of the enigmatic iron-based superconductor FeSe, whose origin 
remains controversial. We apply an itinerant model previously used to describe the spin-excitation anisotropy as measured by neutron scattering measurements, with magnetic fluctuations included within the RPA approximation. The calculated RIXS cross section exhibits overall agreement with the RIXS data, including the high energy spin-excitation anisotropy.   

\tiny
 \keyFont{ \section{Keywords:} iron-based superconductors, nematic order, FeSe, RIXS theory, magnetic fluctuations. } 
\end{abstract}

\section{Introduction}

Identifying the dominant interaction channels, and pinpointing the correct microscopic origin of preferred electronic ordering tendencies in strongly-correlated materials, constitute a challenge to the theoretical description of materials. This is particularly relevant in systems where spin, charge, orbital, and lattice degrees of freedom all strongly couple with one another. For the iron-based superconductors, the main relevant players are spin-density waves, nematic order, and unconventional superconductivity. In this regard,  iron selenide, FeSe, has played a leading role in recent years since its superconducting phase condenses directly from a nematic state without concomitant broken time-reversal symmetry breaking (magnetic order) at lower temperatures \cite{BoehmerKreisel_review,Coldea_2017_review,Kreisel_review}. In addition, FeSe has been in the spotlight due to its  superconducting transition temperature $T_c$, which is tunable by intercalation, pressure, or dimensional reduction (monolayer FeSe on STO) \cite{Kreisel_review}.

Since FeSe enters an orthorhombic phase below $T_n \sim 90$K it exhibits 90 degree rotational symmetry breaking in all measured quantities (of detwinned crystals). However, from comparisons to theoretical calculations the degree of measured rotational symmetry breaking is much too large to be ascribed solely to the bare electronic structure of the orthorhombic phase. Therefore, several theoretical works have explored the possibility of various interaction-driven feedback effects that enhance the symmetry breaking\cite{Fanfarillo2015,Mukherjee2015,Sprau2017,Kreisel2017,Benfatto2018,Hu2018,Bhattacharyya_PRB2020_1} and strongly influence the shape and orbital content of the Fermi pockets\cite{Kreisel2017,Benfatto2018,Liu2018,Fanfarillo2016,christensen_2020}. A particularly simple theoretical framework which includes such effects is the so-called orbital-selective scenario, where the low-energy self-energy is approximated by orbital-dependent, but energy- and momentum-independent, quasi-particle weight factors\cite{Sprau2017,Kreisel2017,deMedici2014,Kostin2018,Bjornson_2020,Cercellier2019,Biswas2018,Zhou2020}. While this is clearly a crude simplification of the full interacting multi-orbital problem, it was shown to provide overall agreement with a  series of different experiments\cite{Kreisel_review}.

More recently, spectroscopic probes have revealed that the Fermi surface of FeSe is exceedingly anisotropic; it appears to be missing an entire electron pocket at the $Y$-point of the Brillouin zone (BZ), as shown in Fig.~\ref{figFS} \cite{YiPRX2019,Huh2020}. This Fermi surface topology does not naturally arise from DFT band structure calculations, even with additional nematic order added to the description\cite{Kreisel_review}. This finding has reinvigorated the discussion of nematicity and the origin of the large electronic anisotropy in FeSe. For example, the lifting of the $Y$-pocket imposes new constraints on the nature of the nematic order, leading to studies of the importance of $d_{xy}$-orbital contributions \cite{christensen_2020,Jiang_16,Scherer2017,Xing17,Eugenio2018,Morten_2019,Rhodes2021}, and important inter-orbital components in the nematic order\cite{Long2020,Steffensen_2021,Tohyama2021}. The latter were shown recently to arise naturally from longer-range Coulomb interactions\cite{Steffensen_2021}. Additionally, the possible non-existence of the $Y$-pocket has important consequences for superconductivity and the need for anisotropy-enhancing self-energy feedback effects. For example, as shown in Ref. \cite{Steffensen_2021}, the highly anisotropic superconducting gap structure of FeSe follows immediately from standard spin-fluctuation mediated pairing without additional self-energy effects applied to the Fermi surface without any electron pocket at the $Y$-point. This conclusion, however, is mainly a direct consequence of the missing $Y$-pocket itself, and does not eliminate the need for self-energy feedback more generally in the theoretical description of FeSe. This is seen, for example, in theoretical modelling of the neutron response of FeSe, where a prominent momentum anisotropy seems only consistent with calculations incorporating self-energy feedback effects\cite{Steffensen_2021} since the possible lifting of the $Y$ pocket alone only yields a very weak anisotropy of the susceptibility between $(\pi,0)$ and $(0,\pi)$ as also presented in Ref. \cite{Rhodes2021}. 

Therefore, further experiments probing the momentum anisotropy of detwinned FeSe are highly desirable. In this respect, Chen {\it et al.} \cite{Chen2019} succeeded in measuring the inelastic neutron scattering response from a mosaic of single FeSe crystals glued on to BaFe$_2$As$_2$, detwinned at low temperatures by the single domain stripe magnetism of the (uniaxially strained) substrate BaFe$_2$As$_2$ material. This experiment revealed highly anisotropic low-energy ($\lesssim 10$ meV) magnetic fluctuations in (detwinned) FeSe with the main scattering taking place near the $(\pi,0)$ position of the BZ. In the superconducting phase, a similarly momentum-anisotropic resonance peak was additionally identified \cite{Chen2019}. These results can be explained by itinerant models that include self-energy effects that 1) suppress $d_{xy}$ orbital contributions to the spin susceptibility predominantly near ($\pi,\pi$), and 2) favor ($\pi,0$) $d_{yz}$ over ($0,\pi$) $d_{xz}$ orbital contributions in the nematic phase\cite{Kreisel15,kreisel_2018}. Only by allowing for such orbital-selective self-energy effects can a standard RPA-like itinerant scenario be made compatible with the neutron data. We stress that this remains true irrespective of whether or not the $Y$-pocket is present at the Fermi surface.   

Recently, the spin excitations were measured to higher energies in detwinned FeSe by RIXS measurements at the Fe-$L_3$ edge \cite{Lu_RIXS}. The RIXS energy spectra revealed clear dispersive broad spin modes. It was found that the spin-excitation anisotropy, as seen by comparing the scattering cross section along the perpendicular $H$ and $K$ high-symmetry directions, remains to high energies ($\sim 200$ meV). This energy scale is substantially larger than the orbital splitting associated with the nematicity, and as pointed out in Ref. \cite{Lu_RIXS}, the amplitude of the spin-excitation anisotropy in nematic FeSe is comparable to that obtained from the spin-wave anisotropy in the magnetically ordered stripe $(\pi,0)$ phase of BaFe$_2$As$_2$\cite{Lu_RIXS,Lu_PRL2018}. 

The RIXS results for detwinned FeSe provide new testing ground for theories of FeSe. At present the origin of nematicity and the degree of localization and correlation is still being discussed. In particular, theoretical works have both applied models based on fully localized or itinerant electrons, in order to explain the peculiar electronic ordering tendencies of FeSe\cite{Kreisel_review}. Here, we compute the RIXS cross section within an itinerant RPA procedure with nematicity included in the bare band structure\cite{Sprau2017}. The applied RIXS framework is similar to that used in Ref. \cite{Kaneshita_2011} where second order perturbation theory involving the absorption and emission process is used to calculate the RIXS intensity from the generalized spin-susceptibility. The latter is then calculated within a random phase approximation (RPA) where additional reduced coherence of the electronic structure\cite{kreisel_2018} is taken into account. We find that the RIXS cross section as calculated for the fully coherent electronic structure exhibits relatively sharp modes, but remains nearly isotropic when comparing the intensity along the $(\pi,0)$ and $(0,\pi)$ directions, irrespective of whether the $Y$ pocket is present or not at the Fermi level. A strong spin-excitation anisotropy inherent in the sharp paramagnons of the itinerant system can be found if self-energy effects in the nematic state are taken into account. Furthermore, we note that this anisotropy persists to high energies much larger than the energy scale of the nematic order parameter of a few tens of $meV$, similar to the experimental findings in a recent RIXS experiment\cite{Lu_RIXS}. The spin-excitation anisotropy in the theoretical intensity at low energies depends sensitively on the orbital content of the Fermi surface. We discuss implications for our general understanding of magnetic fluctuations and electronic structure of FeSe by comparison to the experimentally determined RIXS data from Ref. \cite{Lu_RIXS}.

\section{Model and Method}

\begin{figure}[tb]
\begin{center}
\includegraphics[width=0.6\linewidth]{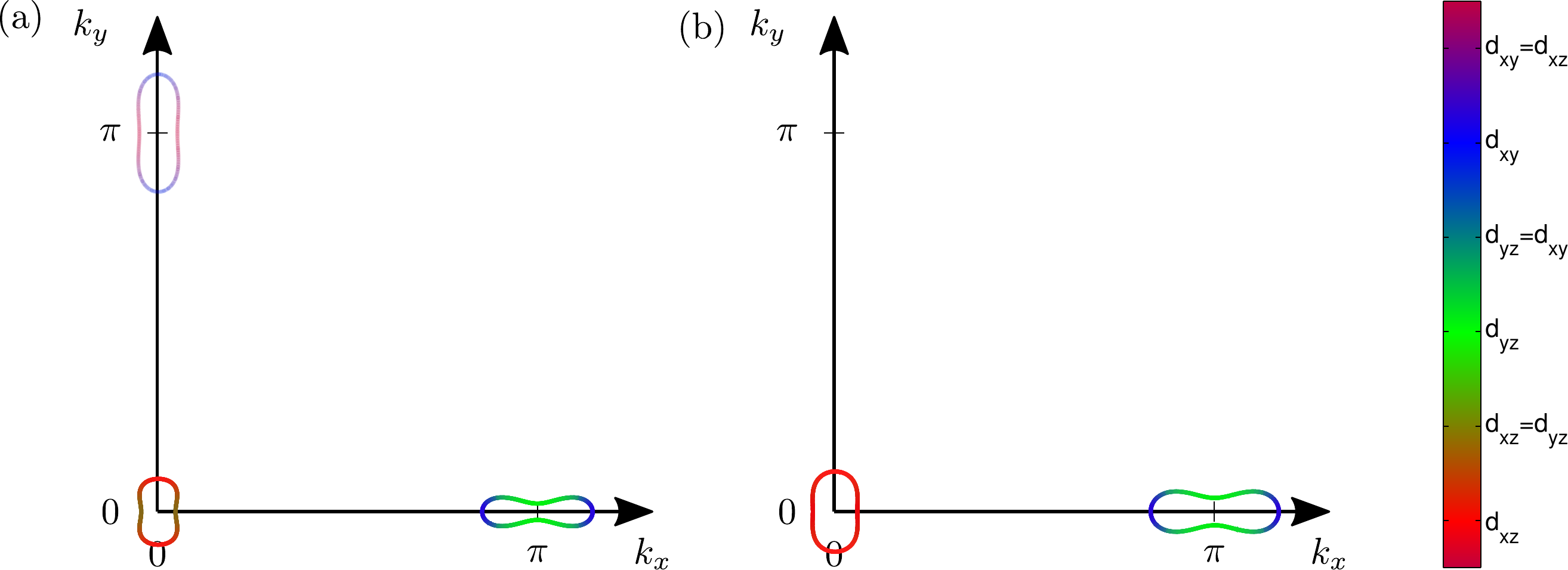}
\end{center}
\caption{Fermi surface of nematic FeSe with orbital content as indicated by colorbar. (a) Model of an electronic structure exhibiting a Fermi surface pocket at the $Y$ point which, however, carries incoherent electronic states (sketched by fading colors)\cite{Sprau2017,Kreisel2017}. Note that the model in Eq. (\ref{eq_tb}) is a three dimensional electronic structure. For the full corresponding Fermi surface we refer to Fig. 1(a) of Ref.\cite{Kreisel2017}. (b) Fermi surface of a model including $d_{xy}$ orbital order as proposed in Ref. \cite{Rhodes2021} where the $Y$ pocket is lifted; similar topology of the Fermi surface was also discussed in \cite{Steffensen_2021} with a different orbital order parameter.\label{figFS}}
\end{figure}

The following calculations are based on a tight-binding parametrization for iron-based superconductors\cite{Eschrig09} with values of the hopping parameters used earlier 
\cite{Sprau2017,Kreisel2017,kreisel_2018}, that closely matches the electronic structure measured in spectroscopic probes. The Fermi surface of this band structure
contains an electron Fermi pocket at the $Y$-point of the BZ,  but its presence is largely irrelevant for the RIXS results discussed below,  compare Fig. 1 (a) of Ref.\cite{Kreisel2017} and Fig. \ref{figFS} for a simplified two dimensional plot of the Fermi surface. Thus, we start from the Hamiltonian
\begin{equation}
H=\sum_{\mathbf{k}\sigma \ell \ell'} t^{\ell \ell'}_{\mathbf{k}} c_{\ell\sigma}^\dagger(\k) c_{\ell'\sigma}(\k),
 \label{eq_tb}
\end{equation}
where $c_{\ell\sigma}^\dagger(\k)$ is the Fourier amplitude of an operator $c_{i\ell\sigma}^\dagger$ that creates an electron in  Wannier orbital $\ell$ with spin $\sigma \in \{\uparrow, \downarrow\}$ and $t^{\ell \ell'}_{\mathbf{k}}$ is the Fourier transform of the hoppings connecting the Fe $3d$ orbitals $(d_{xy},d_{x^2-y^2},d_{xz},d_{yz},d_{3z^2-r^2})$. This term includes an on-site spin-orbit coupling of  type $S^zL^z$, giving rise to imaginary hopping elements\cite{Kreisel13}, which yield a splitting of the two hole-like bands along the $\Gamma$-$Z$ line in the BZ.
The nematic state at low temperatures is modelled by including an onsite and nearest neighbor bond order term with an energy scale of $\approx 10\,\text{meV}$\cite{Sprau2017,kreisel_2018} arising from Coulomb interactions \cite{Jiang_16,Wu_2016arXiv160302055W,Scherer2017,Steffensen_2021}. While other types of orbital order terms have been proposed in the literature\cite{Rhodes2021,Yu2018}, we do not examine these possibilities in this work.

The Bloch Hamiltonian can be diagonalized by a unitary transformation with the matrix elements $a_\mu^\ell(\k)$, such that it reads
\mbox{$H=\sum_{\mathbf{k}\sigma \mu}\tilde E_\mu(\k) c_{\mu\sigma}^\dagger(\k)c_{\mu\sigma}(\k)$}, where  $\tilde E_\mu(\k)$ are the eigenenergies closely matching the maxima of the spectral function as deduced experimentally\cite{Terashima2014,Audouard2015EPL_Hc2,Watson2015,Watson2016,Sprau2017,Kostin2018}.  $c_{\mu\sigma}^\dagger(\k)$ is the Fourier amplitude of electrons in band $\mu$ and momentum $\k$. Furthermore, we adopt an ansatz for the Green's function in orbital space incorporating correlations via quasiparticle weights $Z_\ell$ in orbital $\ell$,

\begin{align}\label{eq_GF}
\tilde G_{\ell\ell'} (\k,\omega_n)&=\sqrt{Z_\ell  Z_{\ell'}} \;\sum_\mu \frac{a_\mu^\ell(\k) a_\mu^{\ell'*}(\k)}{i \omega_n - \tilde E_\mu(\k)}\notag\\
&=\sqrt{Z_\ell  Z_{\ell'}} \;\sum_\mu a_\mu^\ell(\k) a_\mu^{\ell'*}(\k)\tilde G^\mu(\k,\omega_n).
\end{align}
Here $\tilde G^\mu(\k,\omega_n)=[{i \omega_n - \tilde E_\mu(\k)}]^{-1}$ is the coherent Green's function in band space  which in the paramagnetic state is diagonal in spin space, i.e. proportional to $\delta_{\sigma,\sigma'}$.
This ansatz does {\it not} include the actual incoherent spectral weight, and should therefore only describe the low energy properties of the electronic structure. While the quasiparticle weights  are phenomenological parameters, these can also be calculated e.g. by using fluctuation exchange approach\cite{Bjornson_2020}, or slave-boson methods or dynamical mean field theory\cite{RongYu2012,Georges_rev_2013,deMedici2014,Biermann_review,Yi2017,Guterding2017,Medici_review}, qualitatively giving similar trends for the quasiparticle weights, but in detail yielding different band renormalizations and Fermi surfaces, i.e. exhibiting a low-energy Green's function that is not expected to describe the physical properties accurately at low energies. Here, we adopt the values $\{\sqrt{Z_l}\}=[0.2715,0.9717,0.4048,0.9236,0.5916]$
as used in previous investigations\cite{Sprau2017,Kreisel2017,kreisel_2018}; conclusions are robust as long as the quasiparticle weights are chosen within the range presented in Ref. \cite{kreisel_2018}.

To obtain two-particle responses as measured by a RIXS experiment, we adopt a
standard Hubbard-Hund Hamiltonian  for local interactions
\begin{equation}
	H = {U}\sum_{i,\ell}n_{i\ell\uparrow}n_{i\ell\downarrow}+{U}'\sum_{i,\ell'<\ell}n_{i\ell}n_{i\ell'}
+{J}\sum_{i,\ell'<\ell}\sum_{\sigma,\sigma'}c_{i\ell\sigma}^{\dagger}c_{i\ell'\sigma'}^{\dagger}c_{i\ell\sigma'}c_{i\ell'\sigma}
+{J}'\sum_{i,\ell'\neq\ell}c_{i\ell\uparrow}^{\dagger}c_{i\ell\downarrow}^{\dagger}c_{i\ell'\downarrow}c_{i\ell'\uparrow} , \label{H_int}
\end{equation}
where the parameters ${U}$, ${U}'$, ${J}$, ${J}'$ are given in the notation of Kuroki \textit{et al.} \cite{Kuroki2008}. Imposing spin-rotational invariance, i.e. $U'=U-2J$, $J=J'$, there are only two parameters $U$ and $J/U$ left to specify the interactions which we set to values used previously\cite{kreisel_2018}.  

Within the ansatz of Eq. (\ref{eq_GF}), the paramagnetic orbital susceptibility is given by
\begin{align}
	\tilde\chi_{\ell_1 \ell_2 \ell_3 \ell_4}^0 (q) & = - \sum_{k,\mu,\nu }\tilde M_{\ell_1 \ell_2 \ell_3 \ell_4}^{\mu\nu} (\k,\q) \tilde G^{\mu} (k+q) \tilde G^{\nu} (k),   \label{eqn_supersuscept}
\end{align}
where we have adopted the shorthand $k\equiv (\k,\omega_n)$ and defined the abbreviation
\begin{eqnarray}
	\tilde M_{\ell_1 \ell_2 \ell_3 \ell_4}^{\mu\nu} (\k,\q)& =&\sqrt{Z_{\ell_1}Z_{\ell_2}Z_{\ell_3}Z_{\ell_4}}   
	a_\nu^{\ell_4} (\k) a_\nu^{\ell_2,*} (\k) a_\mu^{\ell_1} (\k+\q) a_\mu^{\ell_3,*} (\k+\q).\nonumber
\end{eqnarray}
After performing the internal frequency summation analytically, we calculate $\tilde\chi_{\ell_1 \ell_2 \ell_3 \ell_4}^0$ by integrating over the full BZ, which is just the susceptibility $\chi_{\ell_1 \ell_2 \ell_3 \ell_4}^0$ of a fully coherent Green's function multiplied by the quasiparticle weights
\begin{equation}
 \tilde \chi_{\ell_1 \ell_2 \ell_3 \ell_4}^0 (\q,\omega)=\sqrt{Z_{\ell_1}Z_{\ell_2}Z_{\ell_3}Z_{\ell_4}}\;\chi_{\ell_1 \ell_2 \ell_3 \ell_4}^0 (\q,\omega).
 \label{eq_susc}
\end{equation}

 Two-particle correlation functions of the interacting system with the interacting Hamiltonian of Eq. (\ref{H_int}) can be calculated in the random-phase approximation (RPA) by summing a subset of diagrams (see, e.g. Ref. \cite{Graser2009}) such that the spin part of the RPA susceptibility, $\tilde\chi_1^{\rm RPA}$, is given by
\begin{align}
\label{eqn:RPA}
 \tilde\chi_{1\,\ell_1\ell_2\ell_3\ell_4}^{\rm RPA} (\q,\omega) &= \left\{ \tilde\chi^0 (\q,\omega) \left[1 -\bar U^s \tilde\chi^0 (\q,\omega) \right]^{-1} \right\}_{\ell_1\ell_2\ell_3\ell_4}.
\end{align}
The interaction matrix $\bar U^s$ in orbital space is composed of linear combinations of $U,U',J,J'$. For its detailed form, we refer to e.g. Ref.~\cite{a_kemper_10}.

\begin{figure}[tb]
\begin{center}
\includegraphics[width=0.6\linewidth]{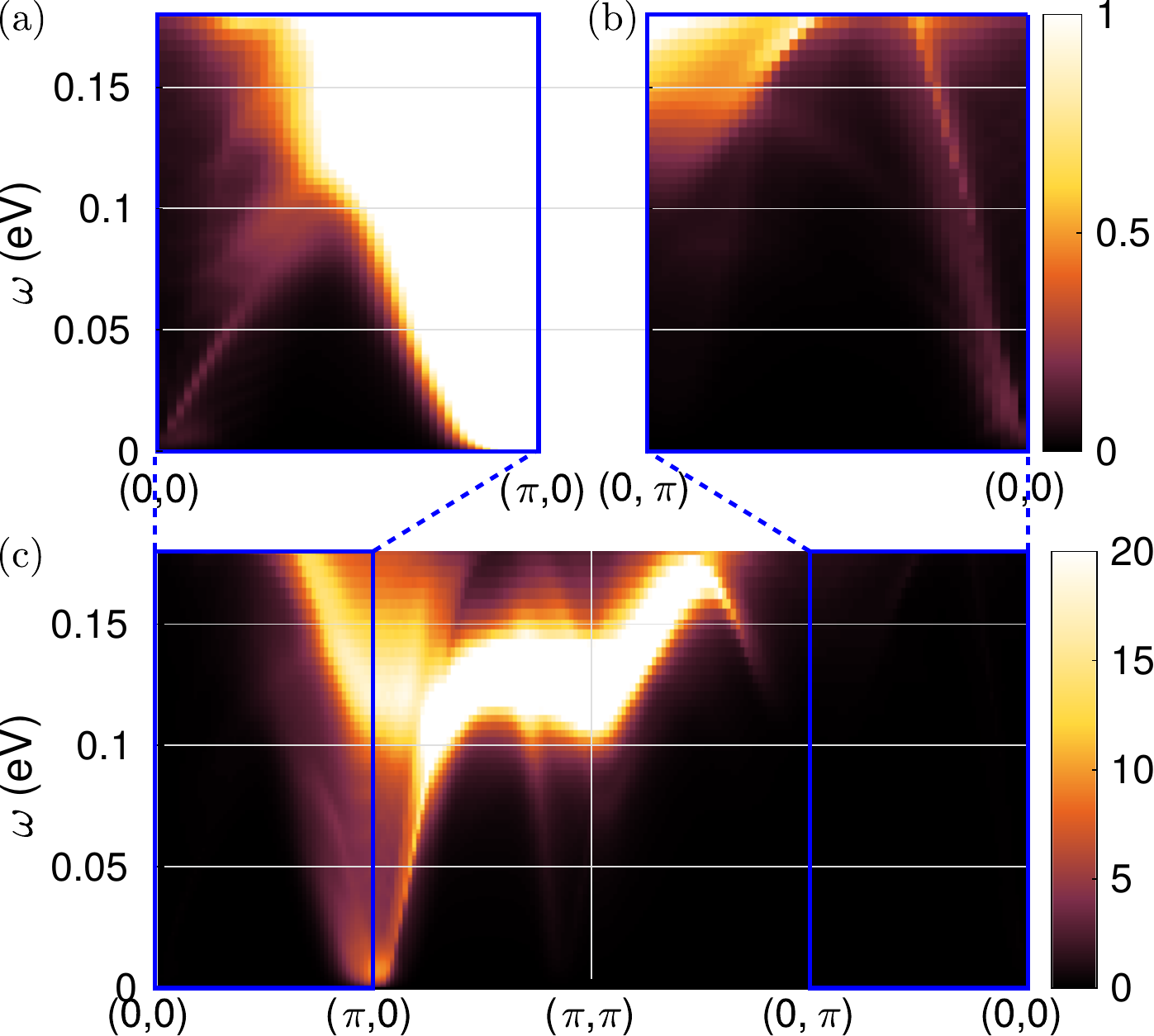}
\end{center}
\caption{Spin susceptibility: $-\mathrm{Im} \chi({\bf q},\omega)$. (a,b) Zoom-in to the details of the spin susceptibility as calculated using the modified RPA approach  for $U=0.57\,\text{eV}$ and $J=U/6$, compare Fig. 9 (c) of Ref. \cite{kreisel_2018}. Close to ${\bf q}=(0,0)$ paramagnon modes are dispersing linearly up as seen towards the $X$ point (a), and towards the $Y$ point (b). The overall intensity close to ${\bf q}=(\pi,0)$ is much larger and exhibits a dispersion with broad maximum around $\omega \geq 0.1 eV$ (c) compared to the relatively sharp paramagnon dispersion close to ${\bf q}=(0,0)$.}\label{fig1}
\end{figure}

The total physical spin susceptibility as, for example, measured in inelastic neutron scattering experiments is then given by the sum
\begin{equation}
	\label{eqn_chisum} \chi (\q,\omega) = \frac 12 \sum_{\ell \ell^\prime} \tilde\chi_{1\;\ell \ell \ell^\prime\ell^\prime}^{\rm RPA} (\q,\omega).
 \end{equation}
For discussion purposes, and to illustrate the differences to the RIXS cross section, we present results for FeSe of this quantity in Fig. \ref{fig1}.v This is the same calculation as in Ref. \cite{kreisel_2018}, but with focus on the small $\bf q$ regions, panels (a,b)).

To calculate the RIXS spectra we follow the approach presented in Ref. \cite{Kaneshita_2011}, where it is calculated as a second-order perturbation from the Kramers-Heisenberg equation in the fast-collision approximation\cite{Ament2011}. 
The transition operator in the dipole approximation, $D_{\mathbf{k}} \approx \sum_{j, j_z,\ell, \sigma,\mathbf{k}'}
c^{j,j_z}_{\ell,\sigma}(\text{\boldmath${\varepsilon}$})
\, c^{\dagger}_{\ell\sigma}(\k) p(\mathbf{k}+\mathbf{k}')_{j,j_z}
+\mathrm{H.c.}$
is written using the fermionic operator of the Fe $2p$ electrons, $
p(\mathbf{k})_{j,j_z}$ for momentum $\mathbf k$ and total angular momentum $j$ and $z$-projection $j_z$. The dipole transition matrix elements $c^{j\,j_z}_{\ell,\sigma}(\text{\boldmath${\varepsilon}$})
=\langle 3d;\ell,\sigma |\mbox{$\text{\boldmath${\varepsilon}$}\cdot\mathbf{r}$}
| 2p;j,j_z \rangle,$ depend on the unit vector of the polarization of the x-rays involved in the process.

Considering the Fe-$L_3$ edge absorption, we restrict to the intermediate $j=3/2$ states of the $2p$ electrons, and calculate the matrix elements assuming wavefunctions with pure hydrogen-like symmetries, i.e. ignoring the deviations of the true Wannier states due to the lower crystal symmetry. The contribution from the radial integration of these wavefunctions will be just a constant (when assuming the same radial dependence for all Fe $3d$ and $2p$ orbitals) while the angular part is given by integrals of trigonometric functions on the unit sphere. Having calculated the matrix elements, one can then obtain the RIXS spectrum from the calculated orbital susceptibility as a sum over internal spin and orbital degrees of freedom via\cite{Kaneshita_2011}
\begin{equation}
I_\mathrm{RIXS}(\mathbf{q},\omega)\propto -\mathrm{Im}\Big\{
\sum_{\{\s_i\}} \sum_{{\ell_1}{\ell_2}{\ell_3}{\ell_4}}
\chi_{{\ell_1}{\ell_2}{\ell_3}{\ell_4}}^{(\s_1,\s'_1)(\s_2,\s'_2)}
(\mathbf{q},\omega)
\Big[\sum_{{j_z,j'_z}}
c^{j\,j_z}_{\ell_1,\s_1}(\text{\boldmath${\varepsilon}$}_\mathrm{o})^*
c^{j\,j_z}_{\ell_2,\s'_1}(\text{\boldmath${\varepsilon}$}_\mathrm{i})
c^{j\,j'_z}_{\ell_3,\s_2}(\text{\boldmath${\varepsilon}$}_\mathrm{i})^* c^{j\,j'_z}_{\ell_4,\s'_2}(\text{\boldmath${\varepsilon}$}_\mathrm{o})\Big]
\Big\},
\label{eq_I-RIXS}
\end{equation}
where $\text{\boldmath${\varepsilon}$}_\mathrm{i}$ and $\text{\boldmath${\varepsilon}$}_\mathrm{o}$ are the polarization vectors of the incoming and outgoing x-rays.
As discussed in Ref. \cite{Kaneshita_2011}, the spin-orbit coupling allows spin-flip processes as mediated by the Clebsh-Gordan coefficients when writing the $2p$ states in the basis for the total angular momentum $j=3/2$. However, since there is no magnetism and we ignore the transverse part of the spin-orbit coupling, the susceptibility turns out to be diagonal, $\chi_{{\ell_1}{\ell_2}{\ell_3}{\ell_4}}^{(\s_1,\s'_1)(\s_2,\s'_2)}
(\mathbf{q},\omega)=\delta_{\s_1,\s_2}\delta_{\s'_1,\s'_2} \tilde\chi_{1\,\ell_1\ell_2\ell_3\ell_4}^{\rm RPA} (\q,\omega)$.

Following the experimental details given in Ref. \cite{Lu_RIXS}, i.e. setting the scattering angle $\beta=50^\circ$, considering the energy of the resonance as $\omega_0=707 eV$, we calculate the polarization vector for incoming $\pi$ polarization as 
\begin{equation}
 \text{\boldmath${\varepsilon}$}_\mathrm{i}= \sin \alpha \mathbf{e}_\parallel+\cos\alpha\mathbf{e}_z,
\end{equation}
where the in plane vector is defined as $\mathbf{e}_\parallel=\mathbf{q}/q$.  
The polarization vectors for the two outgoing polarization directions are
\begin{eqnarray}
  \text{\boldmath${\varepsilon}$}_{\mathrm{o},\sigma}&=& \sin(\alpha+\beta)\mathbf{e}_\parallel+\cos(\alpha+\beta)\mathbf{e}_z,\\
    \text{\boldmath${\varepsilon}$}_{\mathrm{o},\pi}&=& \mathbf{e}_\perp\quad\text{with the perpendicular in-plane vector, i.e. }\,\,\mathbf{e}_\perp\cdot\mathbf{q}=0,
\end{eqnarray}
where the angle $\alpha$ between wavevector $k_\mathrm{i}=\omega_0/(\hbar c)$ of the incoming and the outgoing $k_\mathrm{o}$ x-ray is obtained from solving the equation for momentum conservation along the surface, $q=k_\mathrm{i}\cos\alpha+k_\mathrm{o}\cos(\alpha+\beta)$ for fixed angle $\beta=50^\circ$ and approximating $k_\mathrm{o}\approx k_\mathrm{i}$. Finally, we note that the energy resolution of the RIXS experiment in Ref. \cite{Lu_RIXS} is given as $80\,\text{meV}$. Below, we focus the theory discussion on the as-calculated (non-broadened) computed results.

\section{Results}
\begin{figure}[tb]
\begin{center}
\includegraphics[width=\linewidth]{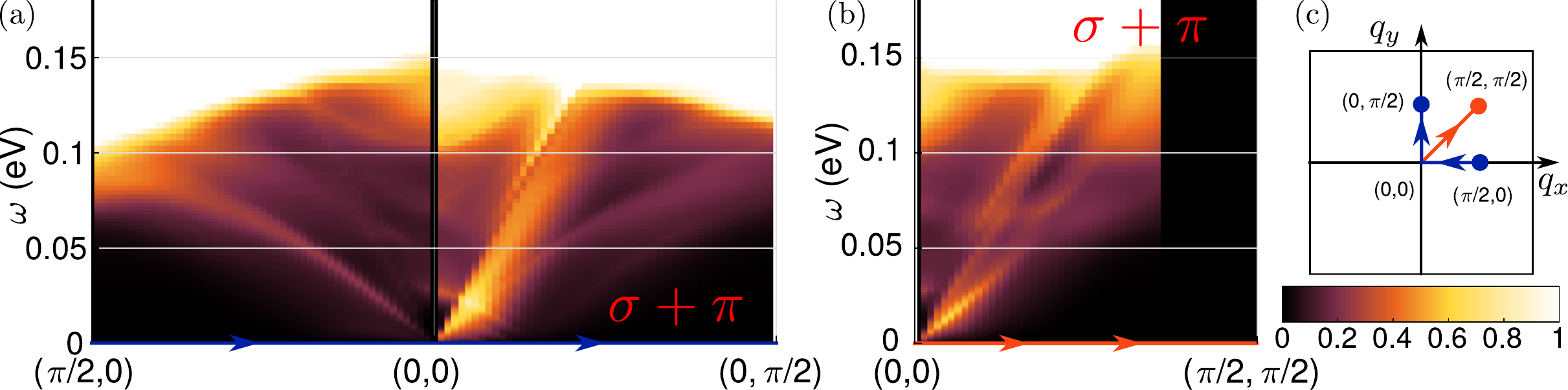}
\end{center}
\caption{RIXS intensity with orbital-selective quasiparticle weight reduction. The RIXS intensity (see common colorbar) at the accessible momentum transfer exhibits sharp paramagnon-like modes towards ${\bf q}=(\pi/2,0)$, while towards ${\bf q}=(0,\pi/2)$ a broad intensity and a much weaker paramagnon mode is visible (a). Along the diagonal in the BZ, there are multiple quasi-sharp paramagnon modes visible (b).  Calculated for $U=0.57\,\text{eV}$, $J=U/6$. Geometry of the paths along the diagonal (orange) and along the coordinate axis (blue) as shown in the other panels (c).}\label{fig2}
\end{figure}

For convenience, and to contrast 
expected intensity measured in an inelastic neutron scattering experiment and a RIXS measurement, we start by presenting the spin susceptibility as obtained from Eq. (\ref{eqn_chisum}) for the case with reduced coherence\cite{kreisel_2018}. In Fig. \ref{fig1} (c) the susceptibility along a high symmetry cut is presented exhibiting large intensity together with a broad dispersive feature close to $(\pi,0)$, and, in contrast, essentially no intensity at $(0,\pi)$. At higher energies, there is also spectral weight close to  $(\pi,\pi)$. Due to the restricted momentum transfer from the photons, RIXS experiments are only able to access the momentum transfer close to $(0,0)$. Therefore the susceptibility in these regions will contribute  to the summation given in Eq. (\ref{eq_I-RIXS}), weighted by the dipole transition matrix elements,  shown in panels (a,b) of Fig.~\ref{fig1} (note different color scale). Already at the level of the (summed) susceptibility, one can see a dispersive and relatively sharp magnetic mode emanating from $(0,0)$ with different slopes along the $q_x$ and $q_y$ directions.

Next, we present our results for the RIXS intensity along high symmetry cuts as detailed in Fig. \ref{fig2} (c), where the sum over the perpendicular polarizations has been performed. It turns out that there is a sharp mode 
along $(0,0)\rightarrow (\pi/2,0)$
that presumably originates from the coherent small $\bf q$-scattering at the $\Gamma$-pocket, which occurs from the $d_{yz}$ orbital component; panel (a). In contrast, there is only a very broad mode along the $(0,\pi/2)$ direction also coming from scattering of the $d_{yz}$ orbital, but at the X-pocket. Scattering contributions from the other orbital components are strongly suppressed due to a reduced quasiparticle weight $Z_\ell<1$. Along the diagonal direction both modes are present, giving rise to two relatively sharp features; panel (b). Note that the black area is due to the mentioned kinematic RIXS constraint, i.e. the respective $\bf q$-vectors cannot be reached.
\begin{figure}[tb]
\begin{center}
\includegraphics[width=0.7\linewidth]{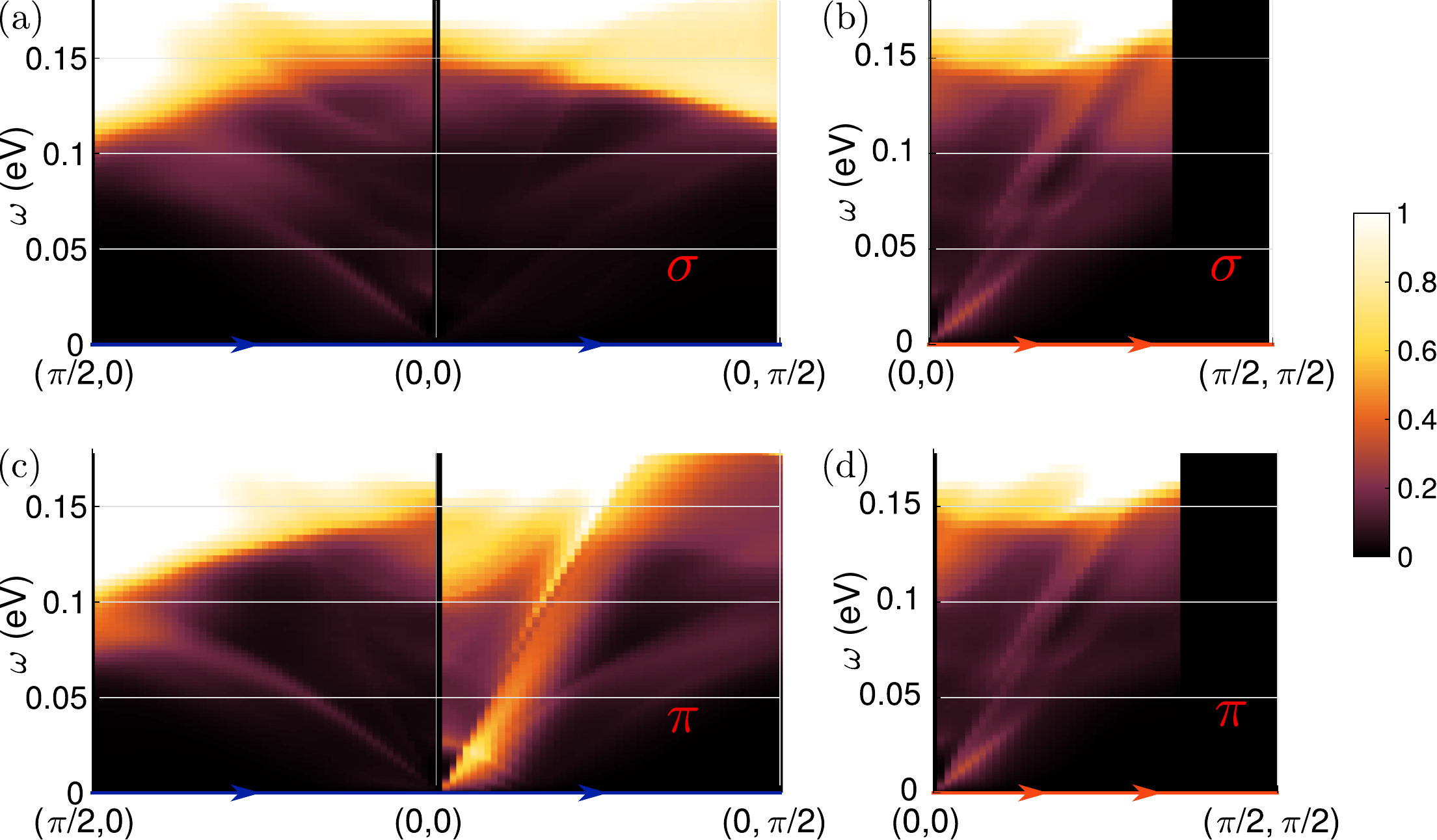}
\end{center}
\caption{Polarization dependence of RIXS intensity. Expected RIXS spectra decomposed in the intensities from $\sigma$ (a,b) and $\pi$ polarization of the outgoing photons (c,d) along the paths defined in Fig. \ref{fig2} (b); $U=0.57\,\text{eV}$, $J=U/6$.}\label{fig3}
\end{figure}

We can disentangle the polarization dependence by looking at each polarization separately. As shown in Fig. \ref{fig3} (a) the $\sigma$ polarization yields a much weaker intensity along the $q_x$ and $q_y$ directions as compared to the $\pi$ polarization, while along the diagonal both polarizations have similar structure and magnitude. One notes also that the broad feature along the $q_y$ cut is only present in the $\pi$ polarization. Indeed, there are strong effects on the anisotropy of the RIXS intensity which are mediated by orbitally selective coherence of the electronic structure, leading to the presence of a sharp mode only along the $q_x$ direction as also detected experimentally; the broad mode along the $q_y$ direction is, however, enhanced due to orbital selectivity. The experimental measurement of the polarization dependence might be able to disentangle scattering contributions from the $\Gamma$- and the $X$-pockets.

\begin{figure}[tb]
\begin{center}
\includegraphics[width=0.7\linewidth]{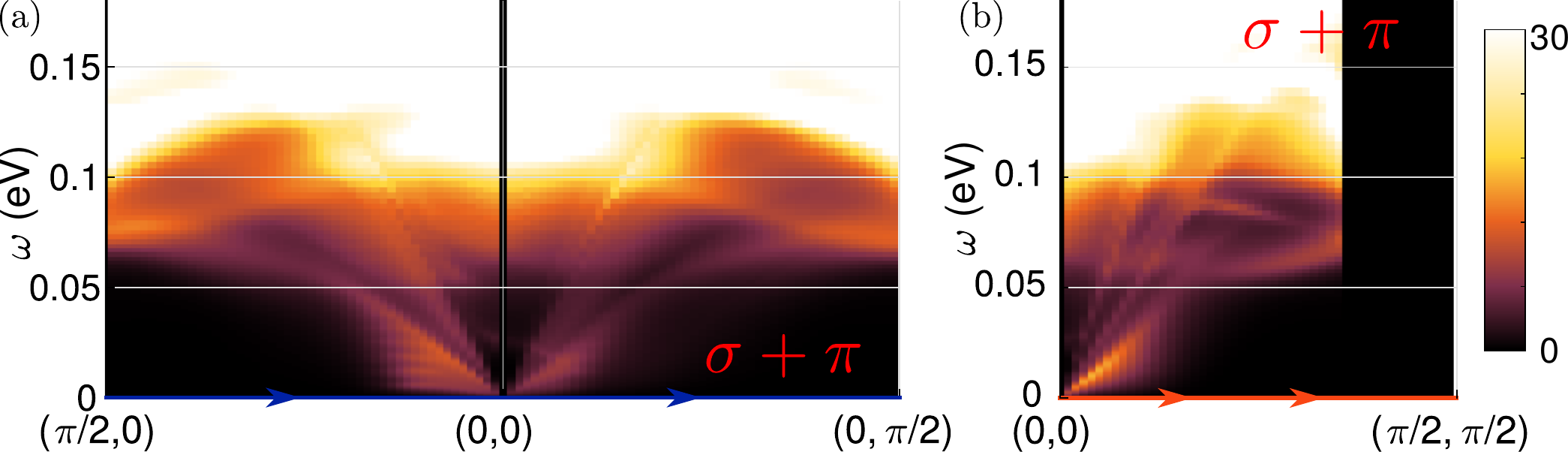}
\end{center}
\caption{RIXS intensity without orbital incoherence. Same as Fig. \ref{fig2}, but calculated using $Z_l=1$ and  by setting $U=0.36\,\text{eV}$, $J=U/6$ as discussed in Ref. \cite{kreisel_2018}.}\label{fig2_2}
\end{figure}

In contrast, a calculation using a fully coherent electronic structure where self-energy corrections are not taken into account, $Z_l=1$, yields a RIXS cross section that is almost isotropic, as shown in Fig. \ref{fig2_2} (a). This result is calculated with the band structure which exhibits a $Y$-pocket at the Fermi level. Except for the very lowest energies $<20 meV$, the same conclusions remain within an electronic structure where the $Y$-pocket has been lifted by nematic order from nearest neighbor Coulomb interactions\cite{Steffensen_2021}, or from including $d_{xy}$ nematicity\cite{Rhodes2021}. The reasons for this almost isotropic result are similar to the spin susceptibility discussion \cite{Kreisel2017}, whereby the missing $Y$-pocket only reduces scattering contributions at very low energies corresponding to the nematic energy scale, while interband contributions and scattering at larger energies are almost identical to the ones from a model with a $Y$-pocket present at the Fermi level.

\begin{figure}[tb]
\begin{center}
\includegraphics[width=0.7\linewidth]{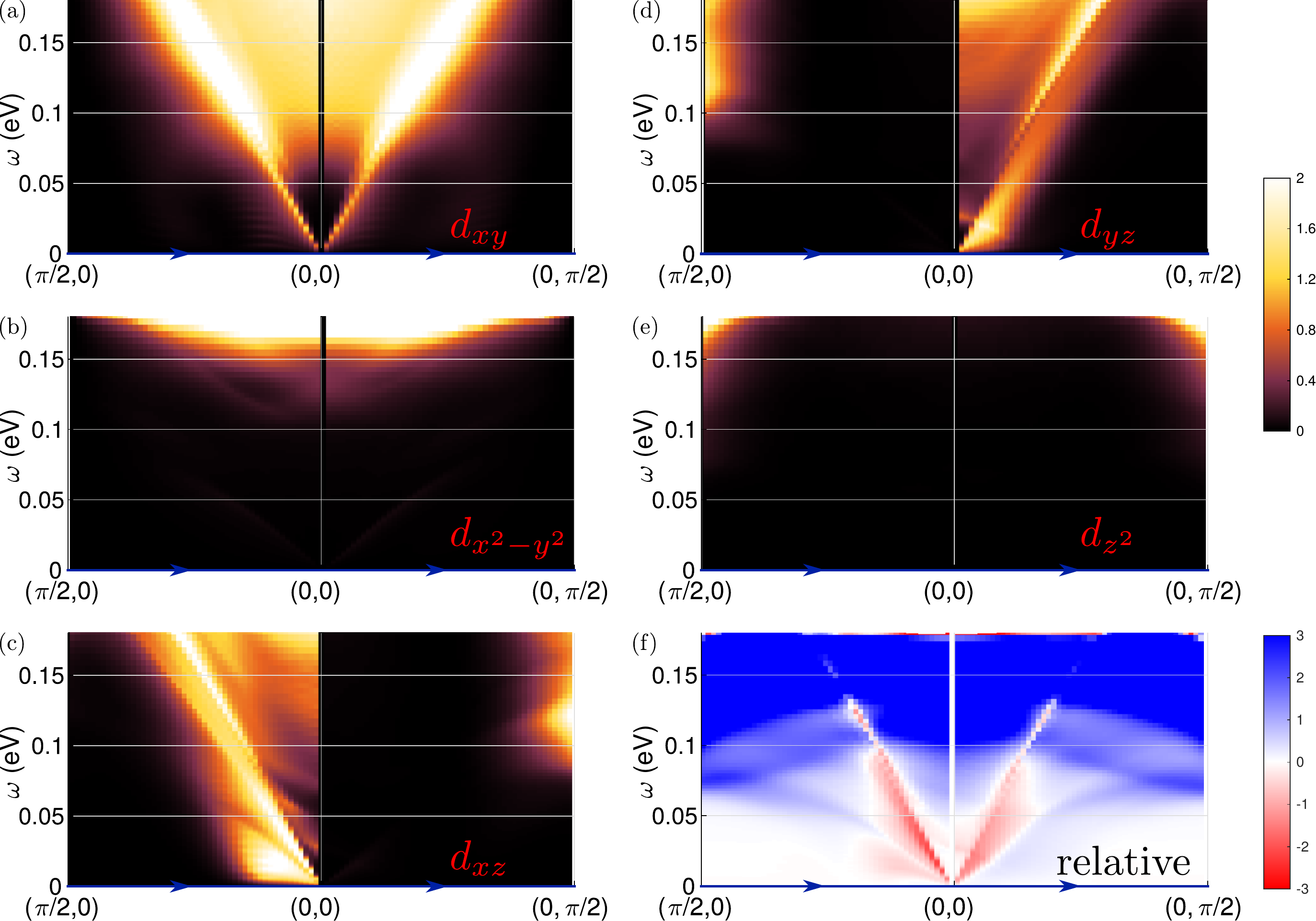}
\end{center}
\caption{RIXS intensity from single orbital components. Intensity calculated by setting $Z_l=0$, except for one orbital component, where $Z_{l^*}=1$ (a-e), i.e. in the sum in Eq. (\ref{eq_I-RIXS}) only the fully diagonal components of the susceptibility contribute. The off-diagonal contributions are sizeable as showin in panel (f) where the results of panels (a-e) multiplied by the number of orbitals are subtracted from the full RIXS intensity as shown in Fig. \ref{fig2_2} (a).}\label{fig2_3}
\end{figure}

To complete our understanding of the origin of the different spectral features in the RIXS intensity, we present in Fig. \ref{fig2_3} a separation of the intensities in orbitally diagonal components, i.e. considering in the sum of Eq. (\ref{eq_I-RIXS}) only the terms with $\ell_1=\ell_2=\ell_3=\ell_4$, panels (a-e), and extracting the off-diagonal contributions by subtracting diagonal components from the full intensity for the case of $Z_l=1$. One clearly sees that the $d_{x^2-y^2}$ and the $d_{z^2}$ orbitals do not contribute to the intensity in the energy range at all. The $d_{xy}$ orbital yields an almost isotropic contribution along the $q_x$ and the $q_y$ cuts which, however, should be suppressed given the correlated nature of that orbital. Finally, the $d_{yz}$ orbital contributes to a branch along $q_y$ from scattering within the $X$-pocket, while the $d_{xz}$ orbital contributes with a slightly larger intensity along the $q_x$ direction. Again, these conclusions remain similar for models without the $Y$-pocket with the exception of reduced weight from the $d_{xz}$ orbital at and below the nematic energy scale of $\approx 20$ meV (not shown).
Panel (f) of Fig. \ref{fig2_3} demonstrates that the orbitally off-diagonal contributions are quite sizeable; we also note that different from the susceptibility extracted from
inelastic neutron scattering experiments, the sum in Eq. (\ref{eq_I-RIXS}) contains elements of the susceptibility tensor with all four orbitals being different. Interestingly, the sharp dispersive mode along $q_x$ as seen in Fig. \ref{fig2} (a) and also Fig. \ref{fig3} (a,c) does not originate from orbitally diagonal contributions, but rather appears as blue mode in the subtracted intensity of Fig. \ref{fig2_3} (f). Hence the orbitally off-diagonal contributions give rise to this intensity, and it is less affected by reduced coherence and therefore more visible in Fig. \ref{fig2} (a) compared to Fig. \ref{fig2_2} (a).

Since the RIXS experiment is kinematically constrained to momentum space close to $(0,0)$ the dispersive modes are less affected by the particular choice of the bare interaction Eq. (\ref{H_int}), i.e. no shift of intensity to lower energy is visible as the magnetic instability is approached, $U\rightarrow U_c$. This is unlike the dispersive modes close to $(\pi,0)$ or $(\pi,\pi)$ whose bandwidth is strongly governed by the denominator in the RPA equation for the susceptibility, Eq. (\ref{eqn:RPA}), i.e. the spectral position of the high energy weight presented in Fig. \ref{fig1}(c) is sensitive to the value of the bare interaction $U$.

In Ref. \cite{Lu_RIXS} the RIXS data was analyzed in terms of a phenomenological model where the RIXS spectra were fitted to a general damped harmonic oscillator model, and discussed in terms of an anisotropic Heisenberg Hamiltonian. In addition, it was concluded without explicit calculations that itinerant models are at variance with the RIXS data due to expected Landau-damped high-energy excitations. The current calculations invalidates this argumentation since we find highly dispersive magnetic excitations persisting to high energies. Indeed, the sharp dispersive mode is visibly strongest along the $q_x$ direction, see Fig. \ref{fig2} (a). In general, we find a spin-excitation anisotropy with larger intensity along the $q_x$-directions, similar to experiments\cite{Lu_RIXS}. At the lowest energies, however, the current band structure produces a larger intensity in the $q_y$ direction; a property which is not seen experimentally\cite{Lu_RIXS}. The reason for this discrepancy is the "boosted" $d_{zy}$ orbital due to the particular choice of quasiparticle weight factors. This hints at more $d_{xz}$-orbital content present at the Fermi level than included in the present modelling. 

\section{Summary and Conclusions}

We have provided a microscopic calculation of the RIXS and neutron response relevant for nematic FeSe. The model is based on itinerant electrons with additional interaction-generated self-energy effects, crudely approximated by simple energy- and momentum independent quasi-particle weight factors. This approach offers a  consistent picture of spin fluctuations as detected in inelastic neutron scattering and the recent RIXS experiments, in addition to other experiments, without further tuning of parameters. Specifically, the calculations yield overall agreement with the momentum and energy structure of the low-energy modes, and their momentum anisotropy. We have also discussed quantitative discrepancies between the current calculation, and the recent RIXS measurement by Lu {\it et al.}\cite{Lu_RIXS}. The microscopic calculation allowed us to explore orbital- and band-dependence of the RIXS scattering cross section, revealing 1) an insensitivity of the RIXS spin-excitation anisotropy response to the presence or absence of a $Y$-pocket at the Fermi level, and 2) a sensitivity of the low-energy anisotropy to the detailed balance of $d_{xz}$- and $d_{yx}$-orbital content present on the $\Gamma$- and $X$-pockets of the Fermi surface.

While the RPA approach to itinerant spin excitations is expected to break down at sufficiently high energies, where exactly this occurs is not clear; the crossover to a more localized description is expected in the range of 100s of meV.  Here we have shown that for intermediate energies of up to $\sim$ 150 meV this approach appears to reproduce qualitative features, and that well-defined spin excitations are not overdamped by electron-hole scattering.  Of course the theory is not complete in the sense that the quasiparticle weights are not derived properly from a self-energy, nor are vertex corrections included.  Nevertheless the current framework appears to be a useful phenomenology to describe the low-energy physics of this unusual material.

\section*{Conflict of Interest Statement}

The authors declare that the research was conducted in the absence of any commercial or financial relationships that could be construed as a potential conflict of interest.

\section*{Author Contributions}

AK performed the calculations, and all authors contributed to writing the manuscript.

\section*{Funding}
B.M.A. acknowledges support from the Independent Research Fund Denmark grant number 8021-00047B.  P.J.H. was supported by the U.S. Department of Energy  under  Grant  No.   DE-FG02-05ER4623.

\section*{Acknowledgments}
We acknowledge useful discussions with A. Kemper and X. Lu.

\section*{Data Availability Statement}
The raw data supporting the conclusion of this article will be
made available by the authors, without undue reservation.

\bibliographystyle{frontiersinHLTH&FPHY} 
\bibliography{references_RIXS}

\end{document}